\documentclass[aps,prb,twocolumn,groupedaddress,showpacs]{revtex4}

\usepackage{graphicx}

\begin{document}

\title{Thermodynamic properties of Pb determined from pressure-dependent critical-field measurements}

\author{N. Suresh$^1$ and J.L. Tallon$^{1,2}$}

\affiliation{$^1$MacDiarmid Institute for Advanced Materials and
Nanotechnology, Industrial Research Ltd., P.O. Box 31310, Lower
Hutt, New Zealand.}

\affiliation{$^2$Victoria University, P.O. Box 600, Wellington,
New Zealand}

\date{\today}

\begin{abstract}
We have carried out extensive low-temperature (1.5 to 10 K)
measurements of the critical field, $H_c$, for the element Pb up to
a pressure of $P=1.2$ GPa. From this data the electronic entropy,
specific heat, thermal expansion coefficient and compressibility is
calculated as a function of temperature, pressure and magnetic
field. The zero-field data is consistent with direct thermodynamic
measurements and the $P$-dependence of $T_c$ and specific heat
coefficient, $\gamma(T,P)$ allows the determination of the
$P$-dependence of the pairing interaction.
\end{abstract}


\maketitle

The thermodynamic and superconducting properties of the metal Pb
have been extensively studied, including specific
heat\cite{Horowitz,Culbert}, isotope effect\cite{Hake,Shaw},
critical field\cite{Decker} and thermal expansion\cite{Andres}. In
addition the effects of pressure on its superconducting properties
have also been well studied\cite{Garfinkel,Brandt,Hansen}.
Subsequently, the use of the pressure-dependent superconducting
transition temperature $T_c(P)$ for Pb as a low-temperature
manometer has also been proposed\cite{Eiling,Bireck,Thomasson}. The
above quoted references are only the early ones and one would assume
that the thermodynamic properties of this material have been
comprehensively investigated.

In the course of investigations into the pressure dependence of the
oxygen isotope effect on $T_c$ for the high-$T_c$ superconductor
YBa$_2$Cu$_4$O$_8$ we used the metal Pb as an internal pressure
calibrant. Using a clamp cell in a SQUID magnetometer we tracked the
pressure using the reported pressure dependence of $T_c$ for
Pb\cite{Eiling}. Despite the many above-noted early studies on Pb we
discovered that the pressure, $P$, and temperature dependence of the
critical field, $H_c$, for Pb has not been reported except at very
low pressures ($\leq$ 0.03 GPa)\cite{Garfinkel}. We therefore
carried out a study of $H_c(T,P)$ from 1.5 to 10 K and up to 1.2 GPa
from which we have derived a full set of thermodynamic parameters.
This paper presents a summary of these measurements and the deduced
values of the $T$- and $P$- dependence of the electronic entropy,
$S(T,P)$, electronic specific heat coefficient, $\gamma(T,P)$,
thermal expansion coefficient, $\alpha(T,P)$ and electronic
compressibility, $\kappa(T,P)$.

\begin{figure}
\centerline{\includegraphics*[width=70mm]{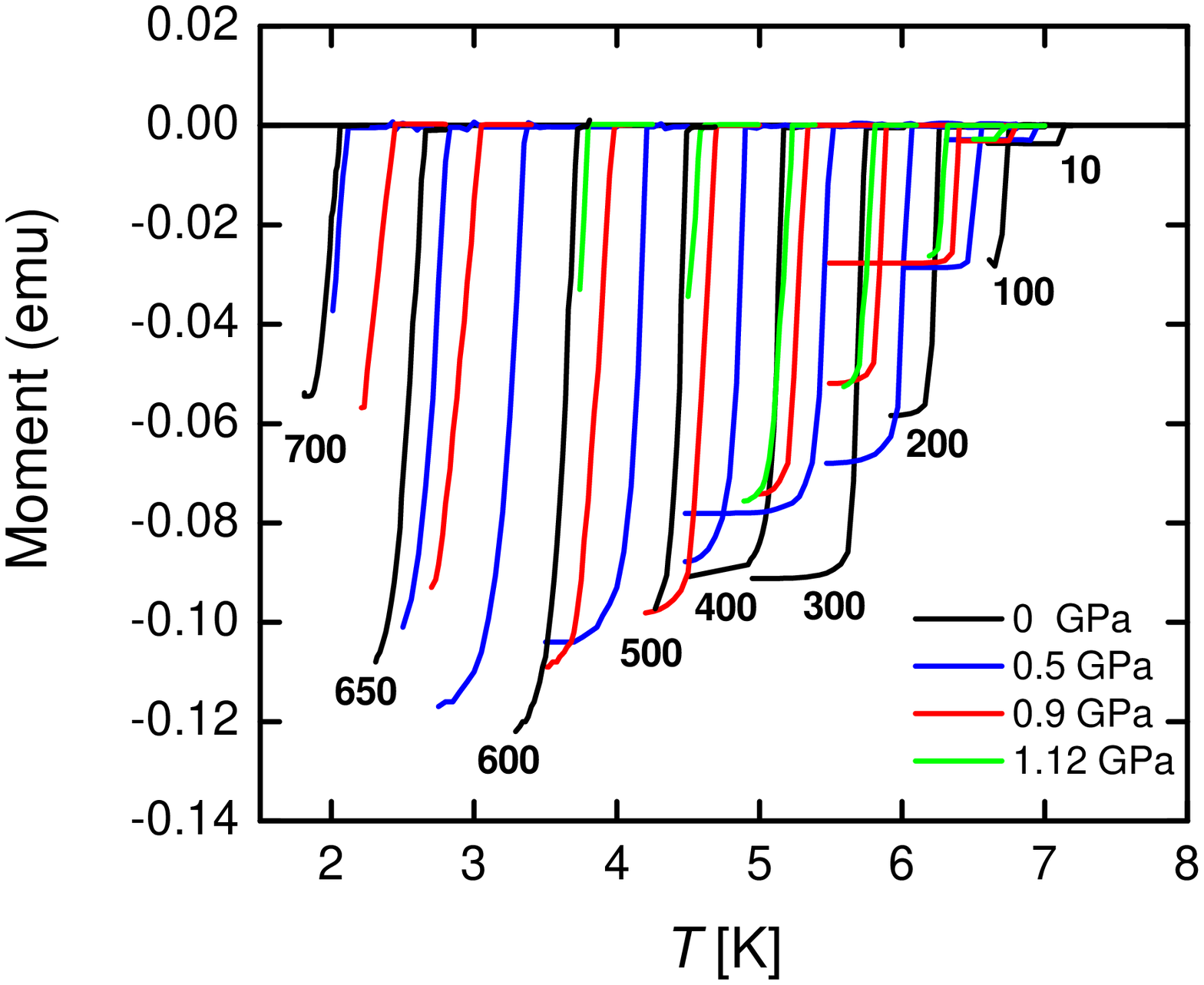}}
\caption{\small (Color) Magnetization curves for the Pb sample
plotted for various applied fields (annotated in Gauss as 10, 100,
200 etc) and four different pressures (0, 0.5, 0.9 and 1.15 GPa).}
\end{figure}
The sample used was a bar of dimensions 1.1 $\times$ 0.93mm cross
section and 5.58mm long, shaped from Puratronic 99.9985\%, 2 mm
diameter Pb wire from Johnson Matthey. Using the formulae of Osborn
the demagnetization factor was calculated as 0.97 and this value was
used throughout. The Pb sample was annealed in vacuum at 300 C for 8
hours to reduce magnetic hysteresis\cite{Decker}. This was loaded in
a 2.67 mm diameter 9 mm long Teflon capsule along with Fluorinert
FC70 and FC77 mixed in 1:1 ratio as a cryogenic hydrostatic pressure
medium. The sample capsule was then placed in a miniature home-built
non-magnetic Be-Cu(Mico Metal 97.75\% Cu, 2\% Be) piston clamp cell
(8.8mm diameter, 65mm length, with cobalt-free tungsten-carbide
pistons (Boride). The pistons of this cell are lightly tapered using
electric-discharge machining\cite{Walker}. To apply pressure the
cell was preloaded before clamping at room temperature using a
laboratory press with calibrated digital pressure gauge (Ashcroft
Model 2089, 0.05\% accuracy). The magnetization measurements were
carried out in a Quantum Design MPMS SQUID magnetometer. The
pressure in the sample was measured from the reported shift in $T_c$
of Pb at zero field\cite{Eiling}. For in-field measurements the cell
was always zero-field cooled to avoid hysteresis error.

Fig. 1 shows the measured $T$-dependence of the magnetic moment,
$M$, up to a field of 700 G and at pressures of 0 GPa (1 bar),
0.5, 0.9 and 1.15 GPa. $T_c$ was determined by the extrapolation
to zero of the steepest slope of $M(T)$. This gives $T_c(H)$ which
is re-plotted in Fig. 2 as $H_c(T,P)$ versus T. The data shows a
progressive decline in both $T_c$ and $H_c(T)$ with increasing
pressure. We will see that the combination of these two pressure
effects allows access to the pressure dependence of the density of
states, $N(E_F)$, at the Fermi level, $E_F$ and of the pairing
interaction.

\begin{figure}
\centerline{\includegraphics*[width=75mm]{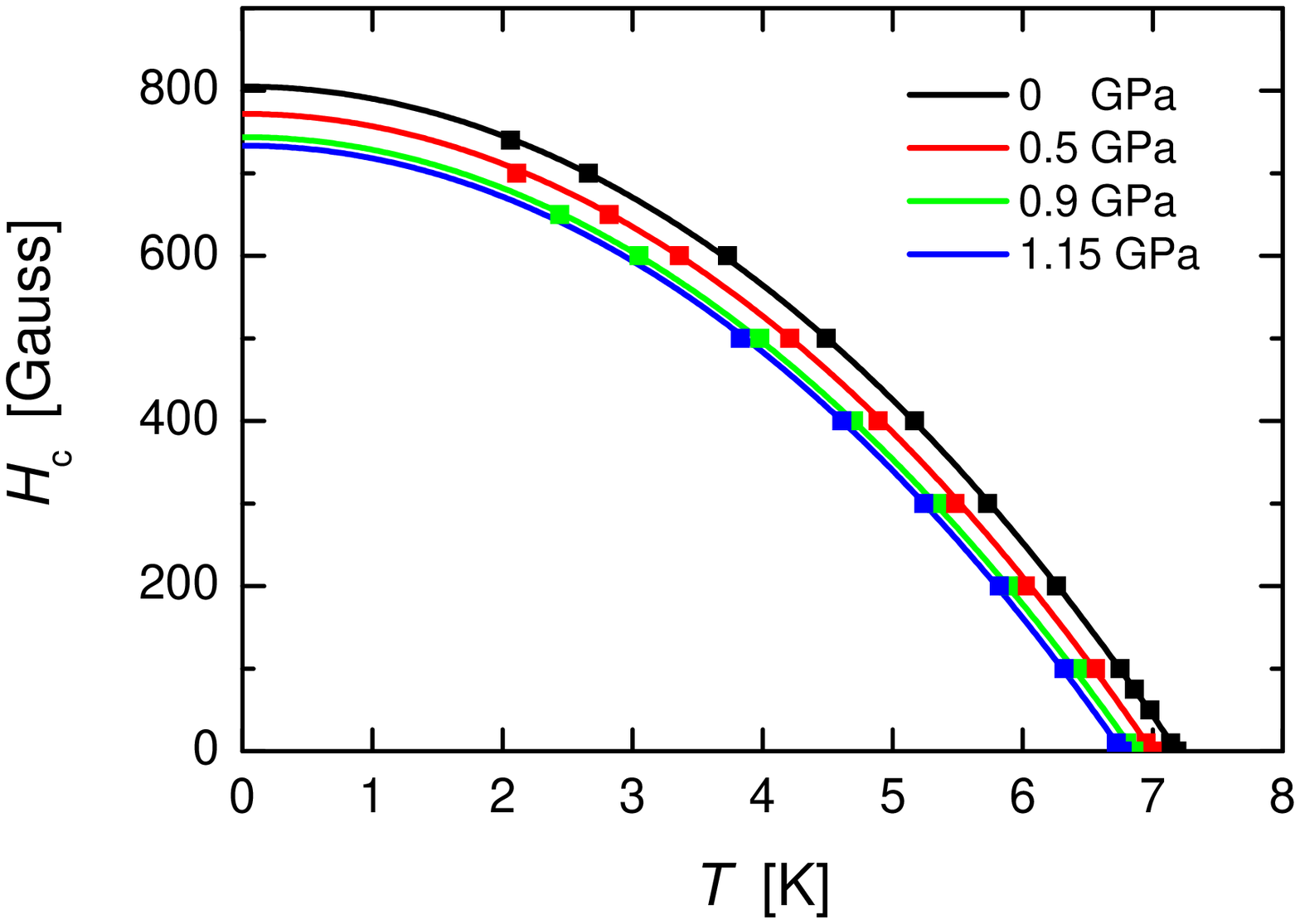}}
\caption{\small (Color) The $T$-dependence of the critical field,
$H_c$ for pressures of 0, 0.5, 0.9 and 1.15 GPa. Data points are
measured data from Fig. 1, curves are fits using eq. 2.}
\end{figure}

To fit this data we expand $H_c(T)$ as a power series in $T$. The
first-order term must be zero otherwise $S_s\neq0$ at $T=0$ and the
third-order term must also be zero to avoid a negative $\gamma$ at
low $T$. It is usual, therefore, to adopt an even polynomial of the
form\cite{Decker}
\begin{equation}
\ H_c(T,P) = H_{c0}(P) + H_{c1}(P)\times T^2 + H_{c2}(P)\times T^4
\end{equation}

\noindent where, as indicated, the coefficients are each an
independent function of pressure. As we will see, $H_c$ cannot be
linear in $T$ because of the third law requirement that the entropy,
$S \rightarrow 0$ as $T \rightarrow 0$. It turns out that the
coefficients are not independent and it is common to adopt the
``similitude principle" which separates the $P$- and $T$-dependent
terms as follows\cite{Garfinkel}
\begin{eqnarray}
\nonumber H_c(T,P) & = & H_{c0}(P)\times f(t)
\\ & = & H_{c0}(P)\times\{1 - \alpha\times t^2 + (1-\alpha)\times t^4\}
\end{eqnarray}

\noindent where $t = T/T_c$. A free fit of eq. (1) to the data
consistently yielded $H_{c2}/H_{c0}$ very close in value to $1 -
H_{c1}/H_{c0}$, consistent with eq. (2). We therefore proceeded to
fit the data with eq. (2) for all subsequent analysis. The fits are
shown by the solid curves through the data shown in Fig. 2. We
obtain $H_{c0}$ values of 804.08, 770.88, 739.99 and 729.01 Gauss
for $P =$ 0, 0.5, 0.9 and 1.15 GPa, respectively. A linear fit
yields
\begin{equation}
\ H_{c0}(P) = (803.72 - 67.028P) \text{  Gauss}
\end{equation}

This now allows the scaling of all the data by plotting
$H_c(T,P)/H_{c0}(P)$ versus $t^2$ as shown in Fig. 3. The small
departure from linearity in Fig. 3 reflects the small quartic term
($\alpha = 0.95475$; $1-\alpha = 0.04525$) and the scaled data in
Fig. 3 allows a global fit to determine $\alpha$ that provides an
overall consistency in the data fits that is necessary when second
derivatives of $H_c(T,P)$ are used to determine the electronic
specific heat coefficient, thermal expansion coefficient or
compressibility.

\begin{figure}
\centerline{\includegraphics*[width=75mm]{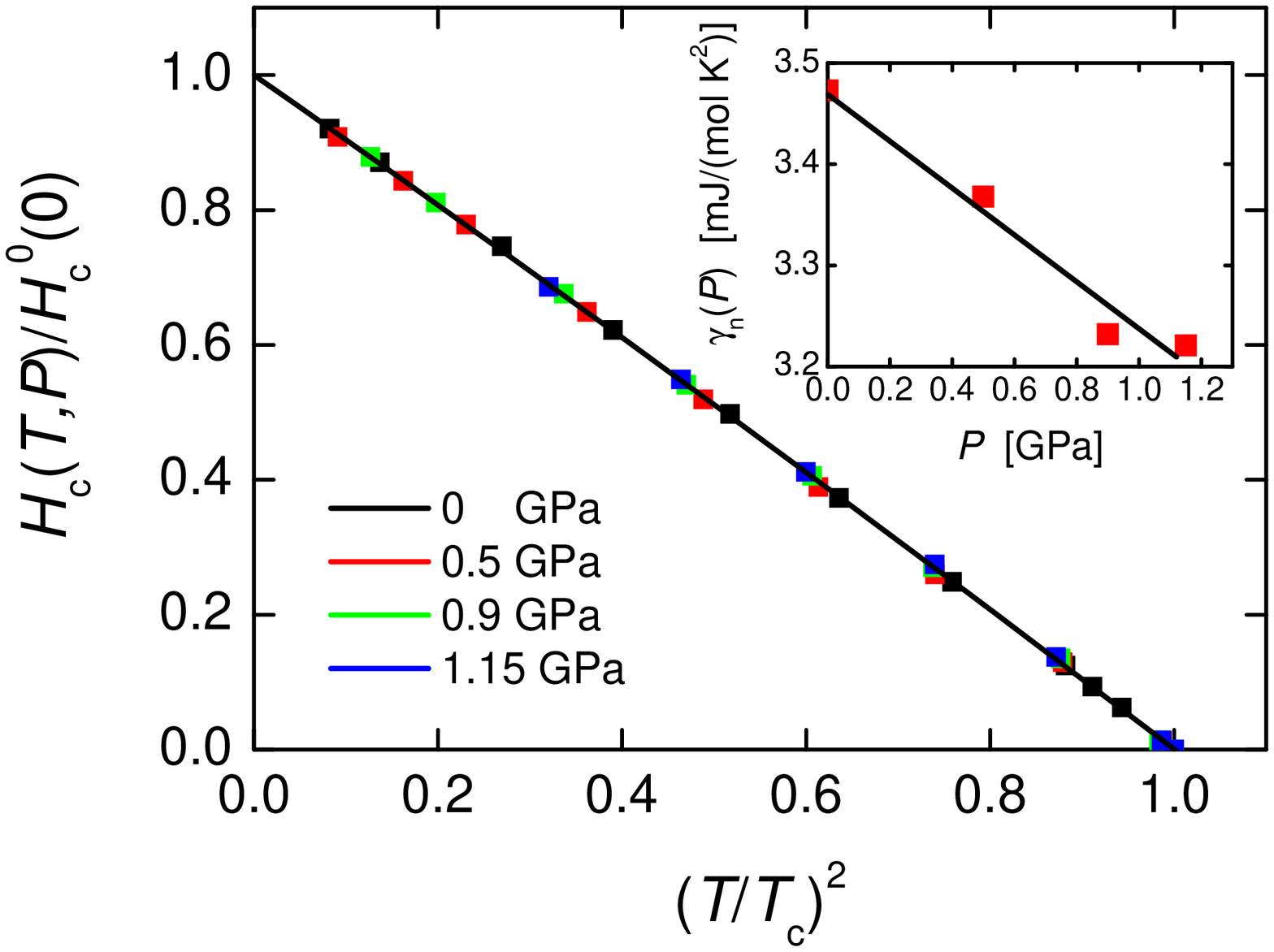}}
\caption{\small (Color) Scaling of the $T$-dependent critical field,
showing $H_c(T,P)/H_c(T=0,P)$ versus $t^2 = (T/T_c)^2$ for pressures
of 0, 0.5, 0.9 and 1.15 GPa. The solid curve is the function $f(T) =
1 - 0.95475t^2 - 0.04525t^4$. Inset: $\gamma_n$ calculated from eqn.
(6).}
\end{figure}

These parameters may be determined as follows. We restrict our
thermodynamic parameters to the electronic contribution, thus
ignoring the lattice contribution to the specific heat or thermal
expansion. The difference in free energy between the normal and
superconducting states at the same $T$ and $P$ is given by
\begin{equation}
\ G_s(T,P) = G_n(T,P) - {1\over2}\mu_o H_c(T,P)^2\times V_M
\end{equation}

\noindent where $\mu_o$ is the permeability of free space and
$V_M$ is the molar volume. As shown by Shoenberg\cite{Shoenberg}
the thermodynamic parameters are obtained by differentiation with
respect to $T$ and $P$, giving
\begin{eqnarray}
\nonumber  S_s(T,P) & = & S_n(T,P) + \mu_o H_cV_M(\partial
H_c/\partial T)_P
\\ \nonumber \gamma_s & = & \gamma_n + \mu_o V_M (\partial H_c/\partial T)_P^2
\\ \nonumber & & \ \ \ \ - \ \mu_o H_cV_M (\partial^2 H_c/\partial T^2)_P
\\ \nonumber V_s & = & V_n - \mu_o H_cV_M(\partial
H_c/\partial P)_T
\\ \nonumber \alpha_s & = & \alpha_n - \mu_o(\partial
H_c/\partial T)_P \times (\partial H_c/\partial P)_T
\\ \nonumber & & \ \ \ \ - \  \mu_oH_c(\partial^2 H_c/\partial T \partial P)
\\ \nonumber \kappa_s & = & \kappa_n + \mu_o(\partial H_c/\partial P)_T^2
\\ & & \ \ \ \ + \  \mu_oH_c(\partial^2 H_c/\partial P^2)_T
\end{eqnarray}

Here terms such as $(1/2)\mu_o H_c^2V_M\alpha_{tot}$ in the
entropy or $\mu_o\kappa_{tot} H_c(\partial H_c/\partial P)_T$ in
the compressibility obtained by differentiation of the molar
volume are ignored as negligible, where $\alpha_{tot}$ is the
total volume coefficient of thermal expansion including lattice
and electronic terms, and similarly for $\kappa_{tot}$.

Taking the first of these expressions, that for the entropy
$S_s(T,P)$, and dividing through by $T$ we note the requirement that
$S_s/T \rightarrow 0$ as $T \rightarrow 0$ due to the opening of a
full superconducting gap in the DOS. At the same time if $\gamma_n$
is constant then $S_n/T$ is just $\gamma_n$. This then imposes a
relationship between $\gamma_n$ and the quadratic term in $H_c(T)$.
In particular,
\begin{equation}
\ \gamma_n = \lim_{T\rightarrow 0} -\mu_oT^{-1}H_cV_M(\partial
H_c/\partial T)_P
\end{equation}

\begin{figure}
\centerline{\includegraphics*[width=75mm]{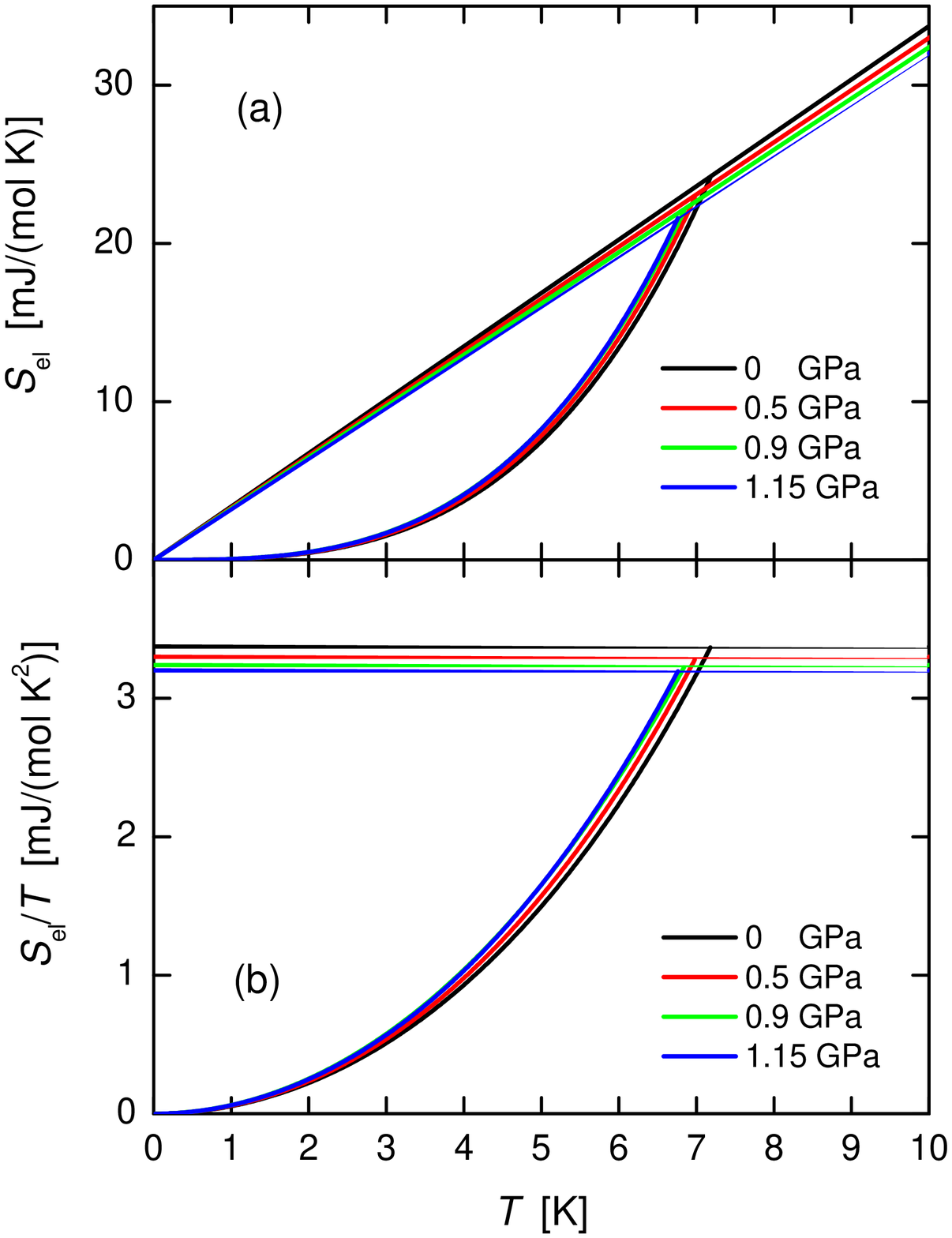}}
\caption{\small (Color) The $T$-dependence of (a) the electronic
entropy in the normal state and superconducting state for pressures
of 0, 0.5, 0.9 and 1.15 GPa; and (b) of $S/T$.}
\end{figure}

In this way $\gamma_n$ is calculated from the fits in Fig. 3 and its
value is plotted in the inset to Fig. 3. The value at ambient
pressure, $\gamma_n = 3.47$ mJ/(mol K$^2$) compares favorably with
the value $\gamma_n = 3.13$ mJ/(mol K$^2$) determined from direct
specific heat measurements\cite{Horowitz}. The pressure dependence
of $\gamma_n$ shown in the inset reveals a dimensionless volume
dependence given by $\partial \ln\gamma_n/\partial \ln V = 3.26 \pm
0.27$ (where we assume a total compressibility of\cite{Waldorf}
$\kappa_0$ = 0.0205 GPa$^{-1}$). There have been various estimates
of this dimensionless parameter ranging from $1.7 \pm 0.5$ from
thermal expansion measurements\cite{White}, $1.8 \pm 0.5$ from
volume expansion at $T_c$\cite{Olsen}, $6.0$ based on pressure
dependence of $H_c$\cite{Garfinkel} and $3.1 \pm 0.8$ in good
agreement with our present value based on more recent measurements
of volume change at $T_c$ by Ott\cite{Ott}. Note that this is
considerably stronger than the value for the free electron gas
$\partial \ln\gamma_n/\partial \ln V = 2/3$.

With these relations in place we proceed to calculate the entropy as
shown in Fig. 4(a). The linear slope in the normal state is just the
above-determined $\gamma_n(P)$. The entropy in both states must
vanish at $T=0$. A more rigorous test of the polynomial fits to the
data is whether $S_s/T$ also vanishes as $T\rightarrow0$ and whether
the data preserves monotonic systematics in this region. The
$T$-dependence of $S/T$ is plotted in Fig. 4(b) and this indeed
shows a quadratic behavior at low $T$ which is perfectly systematic
with increasing pressure and extrapolates to zero as
$T\rightarrow0$. In a finite external field the same entropy curves
are retraced up to the reduced $T_c(H)$ value and then the entropy
jumps discontinuously to the normal-state value, consistent with a
first-order phase transition in magnetic field and second-order when
$H=0$.
\begin{figure}
\centerline{\includegraphics*[width=75mm]{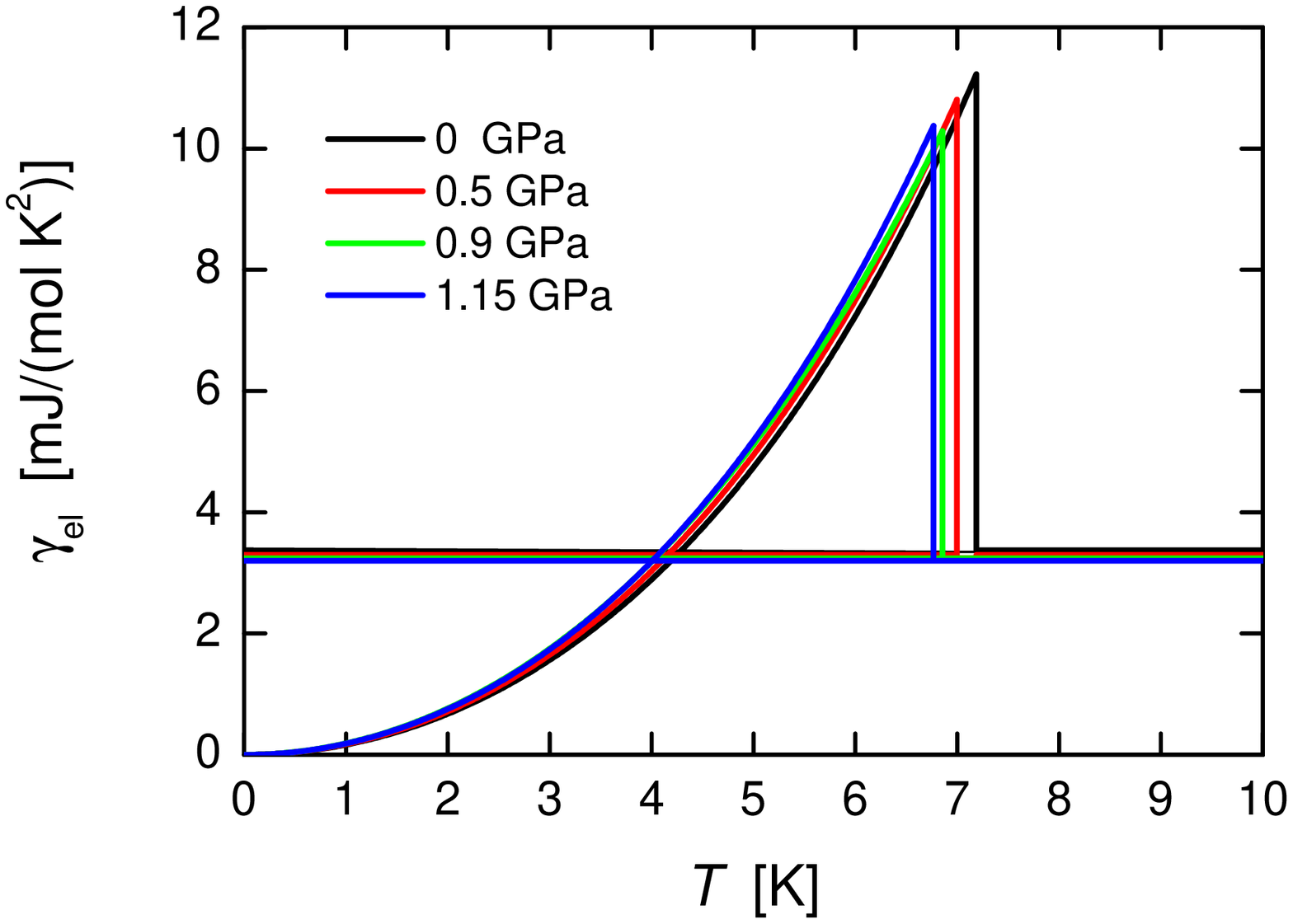}}
\caption{\small (Color) The $T$-dependence of the specific heat
coefficient $\gamma(T,P)$ in the normal and superconducting states
for pressures of 0, 0.5, 0.9 and 1.15 GPa. The small departure from
monotonic systematics in the specific heat jump for 0.9 GPa reflects
the accumulation of errors in the 2$^{nd}$ derivatives.}
\end{figure}

The condensation energy, $U_o$ is determined from
\begin{equation}
\ U_o = \int^{T_c}_0 [S_n(T,P) - S_s(T,P)] dT
\end{equation}

\noindent from which we obtain the following values: $U_o$ = 47,
43.2, 39.8 nd 38.6 mJ/mol for $P$ = 0, 0.5, 0.9 and 1.15 GPa,
respectively. The condensation energy reduces with pressure due to
the twin effects of reduced $T_c$ and reduced $\gamma_n$.

Turning to the specific heat coefficient, this is calculated using
the second equation in eqs. (4) and is shown in Fig. 5 for both the
normal and superconducting states. The entropy balance is evident
from the equal areas above and below the temperature where
$\gamma_s=\gamma_n$, and is confirmed by the fact shown in Fig. 4(b)
that $S_s/T=\gamma_n$ at $T_c$. We find that the jump in
$\gamma(T,P)$ at $T_c$ is $\triangle\gamma_c$ = 7.85, 7.51, 7.05 and
7.17 mJ/(mol K$^2$) for $P$ = 0, 0.5, 0.9 and 1.15 GPa,
respectively. Values of $\triangle\gamma_c/\gamma_n$ are
respectively 2.33, 2.28, 2.18 and 2.24. There is a weak pressure
dependence here in this ratio that is not expected with a simple BCS
picture where $\triangle\gamma_c/\gamma_n = 1.55$. Direct specific
heat measurements at ambient pressure have variously obtained
$\triangle\gamma_c$ = 7.33 $\pm 0.3$ mJ/(mol K$^2$) (Clement and
Quinnel\cite{Clement}), 7.96 $\pm 0.08$ mJ/(mol K$^2$) (Shiffman
{\it et al.}\cite{Shiffman}) and 8.13 $\pm 0.13$ mJ/(mol K$^2$)
(Neighbor {\it et al.}\cite{Neighbor}) in good agreement with our
ambient result. Combining with the measurement by Horowitz {\it et
al}\cite{Horowitz} of $\gamma_n = 3.13$ mJ/(mol K$^2$) gives
$\triangle\gamma_c/\gamma_n$ = 2.34, 2.54 and 2.60, again in good
agreement with our present ambient-pressure result. We note that for
the purely parabolic form $H_c = H_c^o\{1 - t^2\}$ then
$\triangle\gamma_c/\gamma_n$ = 2 exactly, and so the experimentally
observed excess of $\triangle\gamma_c/\gamma_n$ is a direct
indication of the presence of a higher order (quartic) term in the
$T$-dependent critical field.
\begin{figure}
\centerline{\includegraphics*[width=75mm]{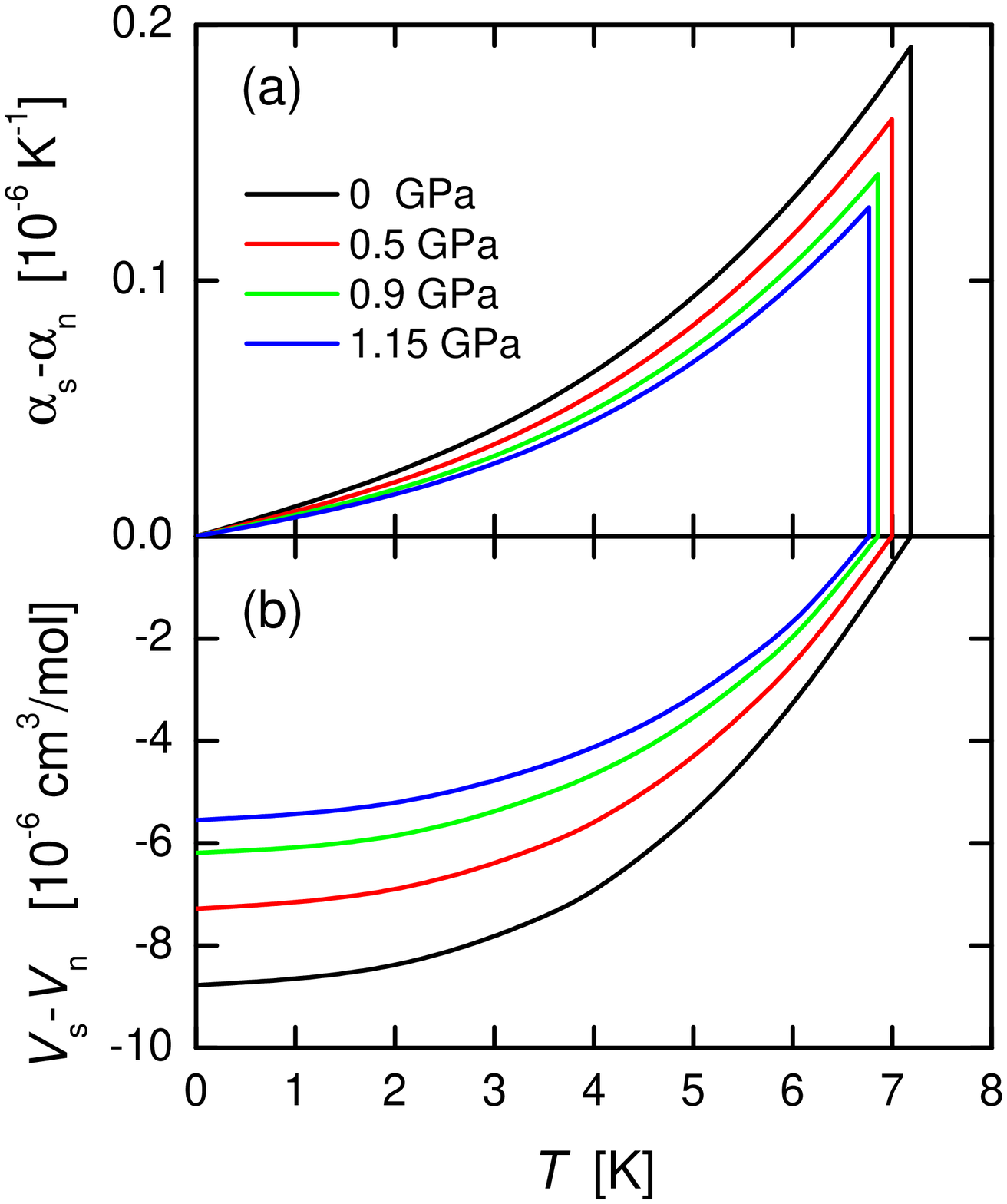}}
\caption{\small (Color) (a) The $T$-dependence of the difference in
volume thermal expansion coefficient between the superconducting and
normal states for pressures of 0, 0.5, 0.9 and 1.15 GPa. (b) the
difference in molar volume for the same pressures.}
\end{figure}

Finally, Fig. 6(a) shows the calculated difference in volume thermal
expansion coefficient between the superconducting and normal states
at each pressure and Fig. 6(b) shows the difference in molar volume.
Given that the absolute molar volume at $T$ = 0 K is 18.26
cm$^3$/mol these changes in volume seem extremely small but they are
readily measureable. The jumps in thermal expansion coefficient at
$T_c$ in zero field are $\triangle\alpha$ = 1.91, 1.63, 1.42 and
1.29 $\times$ 10$^{-7}$ K$^{-1}$ for $P$ = 0, 0.5, 0.9 and 1.15 GPa,
respectively. We omit the calculations of the electronic
compressibility because these involve double derivatives with
respect to pressure and here the errors begin to accumulate
markedly. Nonetheless the discontinuous jumps in isothermal
compressibility may be calculated from the Ehrenfest equation for a
second order transition:
\begin{equation}
\ dT_c/dP = \Delta\kappa/\Delta\alpha
\end{equation}

\noindent Empirically we find that
\begin{equation}
\ T_c(P) = 7.1863 - 0.3847 P + 0.01769 P^2
\end{equation}
\noindent so that the absolute jumps in isothermal compressibility
at $T_c$ in zero field are $\triangle\kappa$ = -7.3, -6.0, -5.0 and
-4.4 $\times$ 10$^{-8}$ GPa$^{-1}$ for $P$ = 0, 0.5, 0.9 and 1.15
GPa, respectively, while the relative jumps are
$\triangle\kappa/\kappa_0$ = -3.56, -2.93, -2.44 and -2.15 ppm.
Direct measurements\cite{Alers} of the ambient-pressure elastic
moduli for Pb in the superconducting state yield
$\triangle\kappa/\kappa_0$ = -4.0 ppm, again in excellent agreement
with our analysis for $P$ = 0 GPa.

We conclude by using the above results to estimate the
pressure-dependence of key parameters in the BCS
model\cite{Meservey}. The condensation energy is
\begin{eqnarray}
\nonumber {1\over2}\mu_0H_{c0}(P) & = & {1\over2}N(0)N_A \Delta_0^2
\\ & = & (47.11 - 7.75P) \text{  mJ/mol}
\end{eqnarray}

\noindent where $P$ is in GPa. The jump in specific heat is given by
\begin{eqnarray}
\nonumber \gamma_s- \gamma_n & = & 10.2 k_B^2 N(0)N_A
\\ & = & (7.85 - 0.68P) \text{  mJ/(mol K}^2).
\end{eqnarray}
\noindent Therefore, we obtain
\begin{eqnarray}
\ N(0) & = & (1.07 - 0.093 P) \text{  states/atom/eV}
\\ \Delta_0 & = & (11.1 - 0.432 P) \text{  K}
\\ 2\Delta_0/k_BT_c & = & 3.09 + 0.037 P.
\end{eqnarray}
\noindent This should be compared with the BCS result
$2\Delta_0/k_BT_c = 3.53$ and the value of $\Delta_0 = 15.76 \pm
0.05$ K observed from tunneling\cite{McMillan} giving
$2\Delta_0/k_BT_c = 4.37$.

If, on the other hand, we determine the DOS from the nearly-free
electron expression\cite{Meservey}
\begin{eqnarray}
\nonumber \gamma_n & = & (2/3)\pi^2k_B^2 N(0)N_A
\\ & = & (3.47 - 0.232P) \text{  mJ/(mol/K}^2).
\end{eqnarray}
\noindent Then we obtain
\begin{eqnarray}
\ N(0) & = & (0.736 - 0.0492 P) \text{  states/atom/eV}
\\ \Delta_0 & = & (13.7 - 0.653 P) \text{  K}
\\ 2\Delta_0/k_BT_c & = & 3.72 + 0.007 P.
\end{eqnarray}
\noindent and the ratio $2\Delta_0/k_BT_c$ is more or less pressure
independent. Finally, using the BCS expression\cite{Meservey} for
$T_c$
\begin{equation}
\ T_c = 0.85\Theta_D \exp(- {{1}\over{N(0)W}})
\end{equation}
\noindent where $\Theta_D$ is the Debye temperature (= 96 K for
Pb\cite{Meservey}) and $W$ is the pairing interaction, then, using
eq. (12)
\begin{equation}
\ W = (0.343 + 0.0167 P) \text{  eV}
\end{equation}
\noindent or using eq. (16)
\begin{equation}
\ W = (0.499 + 0.0143 P) \text{  eV.}
\end{equation}
\noindent Here we have used the Gruneisen coefficient $\gamma_G = -
2.6$ for Pb\cite{Hasegawa}.

In summary, we have measured the pressure dependence of the
superconducting critical field in Pb to 1.15 GPa using a clamp cell
in a SQUID magnetometer. By using thermodynamic identities we have
determined the electronic free energy, entropy, specific heat
coefficient, thermal expansion coefficient and compressibility,
including the jumps in these properties at $T_c$. The calculated
results match rather well the experimentally observed ambient
pressure data, where it is available. The present results allow
calculation of the these parameters as pressure-dependent quantities
and an estimation of the pressure dependence of the density of
states, superconducting energy gap and the pairing interaction.

We acknowledge funding from the New Zealand Marsden Fund, the
International Investment Opportunities Fund and the MacDiarmid
Institute.

\end{document}